\documentclass[grl]{agutex2arxiv}
\usepackage{amsmath,amssymb}
\usepackage{graphicx,color}
\usepackage{lineno}
\usepackage{upgreek}

\setkeys{Gin}{draft=false}

\authorrunninghead{BRANTUT AND MITCHELL}
\titlerunninghead{ASSESSING THERMAL PRESSURISATION}

\authoraddr{N. Brantut and T. M. Mitchell,
  Department of Earth Science,
  University College London, Gower Street, London WC1E 6BT, UK.
  (n.brantut@ucl.ac.uk)}

\begin{document}

\title{Assessing the efficiency of thermal pressurisation using natural pseudotachylyte-bearing rocks}

\authors{Nicolas Brantut, \altaffilmark{1}
  and Thomas M. Mitchell\altaffilmark{1}}

\altaffiltext{1}{Department of Earth Sciences,
  University College London, London, UK.}

\begin{abstract}
  The efficiency of thermal pressurisation as a dynamic weakening mechanism relies on the thermal and hydraulic properties of the rocks forming the fault core. Here, we assess the effectiveness of thermal pressurisation by comparing predictions of temperature rise to field estimates based on pseudotachylyte-bearing rocks. We measure hydraulic and transport properties of a suite of fault rocks (a healed cataclasite, an unhealed breccia and the intact parent rock) from the pseudotachylyte-bearing Gole Larghe fault in the Adamello batholith (Italy), and use them as inputs in numerical simulations of thermal pressurisation. We find that the melting temperature can be reached only if damaged, unhealed rock properties are used. A tenfold increase in permeability, or a fourfold increase in pore compressibility of the intact rock is required to achieve melting. Our results emphasise the importance of damage processes that strongly modify fault rock properties and dynamic weakening processes during earthquake propagation.
\end{abstract}

\begin{article}

  \section{Introduction}


  Weakening of faults during seismic slip is generally considered to be driven by power dissipation and shear heating \citep[][]{ditoro11}. In fluid-saturated fault rocks, shear heating is likely to induce weakening by locally increasing the pore fluid pressure, a process called thermal pressurisation \citep{sibson75,lachenbruch80,mase85,rice06}. In addition to triggerring substantial weakening, thermal pressurisation also maintains faults under relatively low temperature and can prevent frictional melting \citep[e.g.,][]{rempel06,acosta18}. 

  Frictional melting during seismic slip is recognised in exhumed fault zones as pseudotachylytes, which are frozen melt layers containing originally glassy material. Pseudotachylytes are widely recognised as an unequivocal geological record of seismogenic slip along ancient active faults. As they originate from coseismic shear heating, their presence and characteristics also provide key constraints on the shear stresses and power dissipation rates acting during the earthquakes that led to their formation \citep[e.g.,][]{ditoro05}. 
  The very existence of pseudotachylytes also implies that other weakening mechanisms and thermal bufferring processes, such as thermal or chemical (due to devolatilisation reactions) pressurisation, were not efficient enough to prevent melting.

  In fluid-saturated rocks the efficiency of thermal pressurisation as a weakening mechanism (i.e., how quickly and how much strength is reduced from the initial static strength) and as a thermal buffer (limiting how high the temperature can rise during sliding) is governed by a number of physical properties of the fault rock hosting the slip event: thermal capacity, heat diffusivity, poroelastic storage capacity, hydraulic diffusivity, thermal expansivity of the pore space, as well as slip-induced dilation/compaction factors. The effect of thermal pressurisation is typically predicted based on parameter values measured in the laboratory. For instance, \citet{noda05} and \citet{wibberley05} measured permeability, porosity and compressibility on fault gouges sampled along major active faults to predict the slip-weakening behaviour of those faults during earthquakes.  Similarly, \citet{rice06} (followed by a large number of susequent studies) used physical parameters measured in a natural fault gouge \citep[determined by][]{wibberley03} to make general predictions of the characteristic slip weakening distance for upper crustal earthquakes.

  One major challenge when using laboratory-derived data to compute the effects of thermal pressurisation is that the laboratory measurements are conducted on exhumed or subsurface rocks under static conditions that are not representative of the natural conditions prevailing during earthquake slip. In particular, the strong dynamic changes in stress occurring during earthquake propagation are expected to produce significant off-fault damage and therefore strongly increase the permeability of the host rocks \citep[e.g.,][]{mitchell12}. Some ad hoc procedures have been suggested by \citet{rice06} (and subsequently used by \citet{rempel06} and \citet{noda09}) to account for damage in a phenomenological way, by artificially increasing the hydraulic diffusivity of the fault rocks by an arbitrary factor. Despite these efforts, a true assessment of the effect of damage and of the efficiency of thermal pressurisation remains to be achieved.

  Here, we use pseudotachylyte-bearing rocks as a natural thermometer to test predictions from thermal pressurisation, assuming that the rocks are initially fluid-saturated and using laboratory-derived rock properties as model input. We sampled a range of rocks from the Gole Larghe Fault zone (GLFZ), Italy \citep{ditoro04,smith13}, and measured their transport properties as a function of confining pressure in the laboratory. We then use these data in a thermal pressurisation model to predict the peak temperature achieved during a seismic event with comparable magnitude to that which generated the field pseudotachylytes. We test whether these predictions are compatible with the occurrence of pseudotachylytes, and what are the key parameters and unknowns controlling the onset of melting. Overall, we show that using laboratory data from near-fault rocks directly into thermal pressurisation simulations is not compatible with frictional melting, and that some effect of damage (e.g., an increase in storage capacity or permeability) must be accounted for to explain the field observations. Our conclusions highlight the key importance of damage generation and healing during the seismic cycle.

  \section{Laboratory Data on Fault Rocks From GLFZ}

\begin{figure*}
  \centering
  \includegraphics{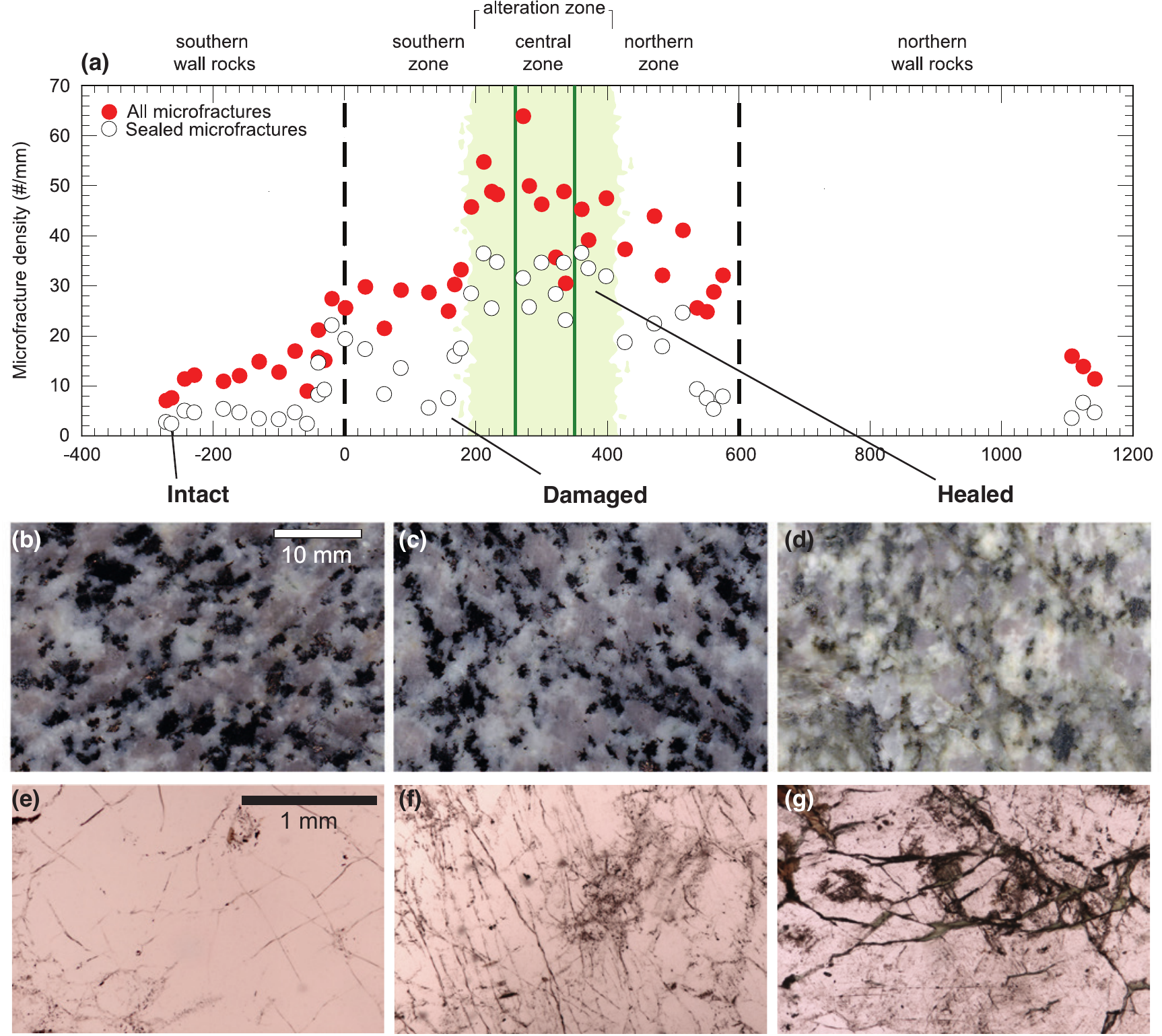}
  \caption{Microfracture density variations acros the GLFZ. (a) Graph of microfracture density vs. distance across the fault zone. Both total microfracture density and the density of sealed microfractures are shown \protect\citep[modified from ][]{smith13}.  (b--d) Images of scanned cut surfaces of the intact, damaged and sealed samples.  Macroscopically both the intact and damaged samples are similar, indicating the majority of damage is microscopic.  The sealed sample illustrates the progressive transformation of brown biotite in the host rock to green chlorite approaching the central area of the GLFZ. (e--g) Optical images of thin sections of intact, damaged and sealed samples, including traces of microfracture damage.}
  \label{fig:field}
\end{figure*}


  The GLFZ is an exhumed strike slip fault in the Adamello batholith in the Italian Southern Alps \citep{ditoro04,smith13}, crosscutting tonalite of the relatively young Val d'Avio-Val di Genova pluton \citep[34--32 Ma;][]{delmoro83}.  Faulting along the GLFZ occurred at around 30 Ma \citep{pennacchioni06}, subsequent to the pluton emplacement and concomitant with faulting along the Tonale Line \citep{stipp04}, at temperatures in the order of 250 to 300$^\circ$C and at a depth of 9 to 11 km.  The GLFZ is a dextral strike-slip fault with a length of around 20 km and a damage zone thickness of around 500 to 600~m \citep{smith13}. Fault damage is characterised by faults with cataclastic cores that are primarily located on preexisting reactivated cooling joints and pseudotachylyte bearing faults.  These structures strike approximately east-west and dip steeply to the south.

  The presence of hydrous phases in the host cataclasites indicates that water must have been present prior to the seismic events leading to the formation of pseudotachylytes. Geochemical analyses by \citet{mittempergher14} suggest that the fluids were likely of metamorphic origin, but the exact pressure conditions are unknown. The chemical alteration within the cataclasites has likely closed the fault core to long term fluid flow, so that any free fluids would have remained trapped at undetermined pressures. Here, we assume that free fluids, pressurised to some degree, were present immediately prior to the pseudotachylyte-forming seismic ruptures. This assumption is discussed is Section \ref{sec:discussion}, but is a realistic one because hydration reactions produce a net volume increase (hence their sealing capability), which has the potential to locally increase the pressure of the remaining pore fluids.

  The damage structure surrounding the GLFZ at both the microscale and macroscale has been investigated previously by \citet{smith13} on fault perpendicular transects (Figure \ref{fig:field}, microscale damage).  Here we chose three representative samples from this study that were (a) intact, (b) damaged, and (c) healed where microfracture damage was previously quantified, and prepared additional samples from the same blocks for measurements of permeability.  

  Microfracture densities are low in the host rock outside the damage zone at $-300$~m, and this is our sample (a) ``intact''.  From the edge of the fault zone, microfracture density increases with increasing proximity to the fault core. We select a representative sample at around $-210$~m of (b) ``damaged'' rock, where open microfracture density is highest.  While microfracture density is highest between two meter-thick cataclastic bands that define the central core zone, microfractures are frequently sealed with K-feldspar, epidote and chlorite, also recognizable in the field by the green colouring of the rocks (Figure \ref{fig:field}c), and this represents our (c) ``healed'' sample.

  We prepared samples from the same oriented blocks collected by \citet{smith13}.  Cylindrical cores with lengths and diameter of up to 75 and 30~mm respectively were prepared, with the core axis perpendicular to the GLFZ plane. The experimental setup consisted of a pressure vessel in which a jacketed sample was subjected to hydrostatic pressure using oil as the confining medium. The pore pressure system, comprising an upstream and a downstream reservoir, was filled with distilled water. Confining pressure and upstream pore pressure were hydraulically servo-controlled \citep[e.g.,][]{duda13}.

  \begin{figure}
    \begin{center}
      \includegraphics{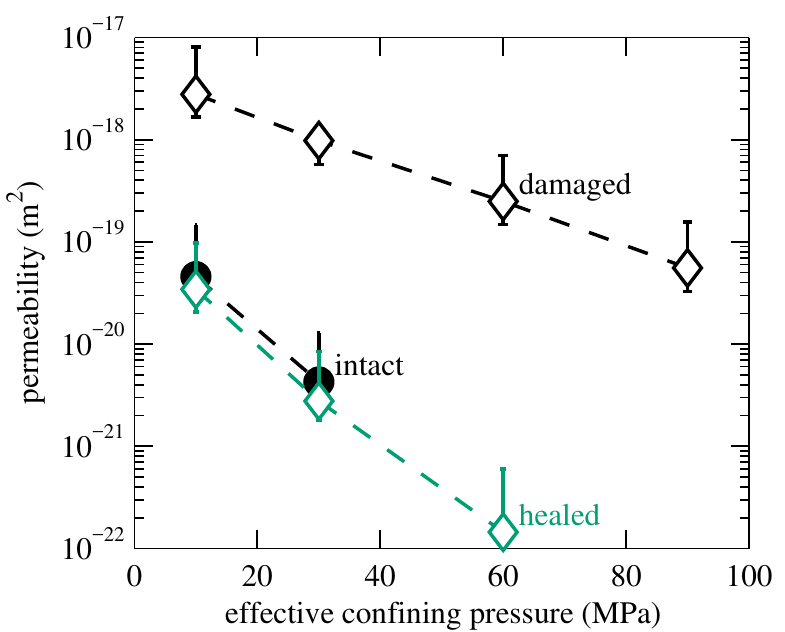}
      \caption{Permeability as a function of effective pressure for three representative samples of the intact, damged-healed and damaged tonalite hosting the pseudotachylytes.}
      \label{fig:kp}
    \end{center}
  \end{figure}

  The permeability of rock samples was measured as a function of effective confining pressure using the method of pore pressure oscillation \cite{fischer92,bernabe06}. The results are shown in Figure \ref{fig:kp}. The permeability of the damaged-healed tonalite is the same as that of the intact one (sampled far from the fault zone), and decreases from around $4\times 10^{-20}$~m$^2$ to $10^{-22}$~m$^2$ for effective pressures increasing from 10 to 60~MPa. The damaged tonalite is significantly more permeable than the intact one, with a permeability ranging from $3\times 10^{-18}$~m$^2$ at 10~MPa effective pressure down to $6\times 10^{-20}$~m$^2$ at 90~MPa effective pressure. In all cases, the permeability is well approximated by a decreasing exponential:
  \begin{linenomath}
    \begin{equation}
      k(P_\mathrm{eff}) = k_0\exp(-\kappa P_\mathrm{eff}),
    \end{equation}
  \end{linenomath}
  where $P_\mathrm{eff}$ is the effective confining pressure, $k_0$ is the nominal permeability value and $\kappa$ is the pressure sensitivity parameter. We find $k_0 = 9\times 10^{-20}$~m$^2$ and $\kappa=1.09\times 10^{-7}$~Pa$^{-1}$ for the intact or damaged-healed tonalite, and $k_0 = 4\times 10^{-18}$~m$^2$ and $\kappa=4.8\times 10^{-8}$~Pa$^{-1}$ for the damaged tonalite.



  The storage capacity of the samples could not be determined accurately across the whole pressure range by the pore pressure oscillation method employed here. However, we obtained useful upper bounds at the lowest pressures. At $10$~MPa effective pressure, in the intact tonalite, a storage capacity $S\approx 9.5\times 10^{-11}$~Pa$^{-1}$ was estimated; in the damaged tonalite, we determined $S\approx 2.5\times 10^{-11}$~Pa$^{-1}$. In the damaged-healed tonalite, the storage capacity could not be determined at all: The inversion method used to extract $S$ from the pore pressure oscillation data for this rock would only output unrealistically low values, less than the compressibility of solid (of the order of $10^{-12}$~Pa$^{-1}$).

  \section{Tests of Temperature Predictions From Thermal Pressurisation}

  \subsection{Governing Equations and Assumptions}

  During rapid shear across a narrow fault core, the temperature ($T$) and fluid pressure ($p$) increase due to shear heating and diffusion across the fault are given by \citep[e.g.,][]{lachenbruch80}:
  \begin{linenomath}
    \begin{align}
      \frac{\partial p}{\partial t} &= \frac{1}{\rho_\mathrm{f}\beta^*}\frac{\partial}{\partial y}\left(\frac{\rho_\mathrm{f}k}{\eta}\frac{\partial p}{\partial y}\right) + \Lambda\frac{\partial T}{\partial t},\label{eq:dpdt}\\
      \frac{\partial T}{\partial t} &= \alpha_\mathrm{th}\frac{\partial^2T}{\partial y^2} + \frac{\tau\dot{\gamma}}{\rho c}, \label{eq:dTdt}
    \end{align}
  \end{linenomath}
  where $t$ is the time, $y$ is the across-fault coordinate, $\tau$ is the shear stress, and $\dot{\gamma}$ is the shear strain rate. A number of material properties appear: $\rho_\mathrm{f}$ and $\eta$ are the fluid density and viscosity, respectively, and $\beta^*$, $k$, $\alpha_\mathrm{th}$ and $\rho c$ are, respectively, the storage capacity, permeability, thermal diffusivity and specific heat capacity of the fluid-saturated rock. The factor $\Lambda$ is given by
  \begin{linenomath}
    \begin{equation}
      \Lambda = \frac{\lambda_\mathrm{f}-\lambda_\mathrm{n}}{\beta_\mathrm{f}+\beta_\mathrm{n}},
    \end{equation}
  \end{linenomath}
  where $\lambda_\mathrm{f}$ and $\lambda_\mathrm{n}$ are the thermal expansivity coefficients of water and of the pore space, respectively, and $\beta_\mathrm{f}$ and $\beta_\mathrm{n}$ are the bulk compressibilities of water and of the pore space, respectively.

  As demonstrated by \citet[][Appendix A]{rice06}, the compressibility of the pore space $\beta_\mathrm{n}$ is not a straightforward material parameter. Assuming elastic fault walls, it is computed as
  \begin{linenomath}
    \begin{equation}
      \label{eq:betan}
\beta_\mathrm{n} = \frac{(\beta_\mathrm{d}-\beta_\mathrm{s})(\beta_\mathrm{d}+r\beta_\mathrm{s})}{n(1+r)\beta_\mathrm{d}}-\beta_\mathrm{s},
    \end{equation}
  \end{linenomath}
where $\beta_\mathrm{d}$ is the drained compressibility of the porous rock, $\beta_\mathrm{s}$ is the compressibility of the solid skeleton, $n$ is the porosity and $r$ is a function of the drained Poisson's ratio of the rock. The drained compressibility is related to the storage capacity as follows:
\begin{linenomath}
  \begin{equation}
    \label{eq:betad}
    \beta_\mathrm{d} = S - n\beta_\mathrm{f} + (1+n)\beta_\mathrm{s}.
  \end{equation}
\end{linenomath}
Using reasonable parameter values of $n=0.01$, $r=1$, $\beta_\mathrm{s}=2\times 10^{-11}$~Pa$^{-1}$ and a conservative bound for the storage capacity of $S=3\times 10^{-11}$~Pa$^{-1}$ for granite \citep{jaeger07}, we determine an estimate of $\beta_\mathrm{n}=2\times 10^{-9}$~Pa$^{-1}$. As a first conservative approximation, we use this value for both the intact and damaged samples, and explore the impact of variations in this parameter later on.

  We assume that the shear strain is homogeneous across a narrow zone of constant thickness $h$, so that $\dot{\gamma}=V/h$, where $V$ is the slip rate during the earthquake. Thermal pressurisation is known to promote strain localisation, unless the shear zone thickness is very small; we therefore choose $h=10$~$\upmu$m, which is equal to, or smaller than, the typical shear zone thickness obtained from spontaneous localisation during thermal pressurisation \citep{rice14,platt14}. This choice is a conservative one in the sense that such a narrow slip zone will tend to promote faster and higher temperature rise during slip than a wider slip zone. 

  We use a realistic, nonconstant slip rate history in the form of a regularised Yoffe function \citep{tinti05}, characterised by a maximum slip, a total duration, and a so-called ``smoothing'' timescale which is a measure of the time between the onset of slip and the peak slip rate. Since pseudotachylytes are observed along small faults with offsets as small as $20$~cm, we choose a maximum slip equal to $20$~cm, a smoothing time of $0.01$~s (so that the peak slip rate is of around $2$~m~s$^{-1}$), and a total duration of $1.02$~s (i.e., typical parameters for a magnitude $M_\mathrm{w}\sim5$ event).

  The shear stress acting on the fault during slip is assumed equal to a constant friction coefficient $f$ multiplied by the effective normal stress $\sigma_\mathrm{n}-p$. The friction coefficient is taken equal to $0.15$, which assumes that high velocity frictional weakening (for instance due to the flash heating) has already occurred during the very early stages of slip \citep[e.g.,][]{brantut16, hayward16}.

The initial temperature of the fault is assumed to be equal to that at $10$~km depth in a relatively hot continental geothermal gradient, which amounts to $T_0 = 300^\circ$C. In the absence of accurate constraints on the fault geometry and stress history, we follow common practice \citep[e.g.,][]{rice06} and assume that the fault normal stress is equal to the lithostatic pressure, that is, $\sigma_\mathrm{n}=280$~MPa. The initial pore pressure is difficult to assess and in the absence of better constraints we choose an approximately hydrostatic fluid pressure condition of $p_0 = 100$~MPa. This hypothesis is the most common is thermal pressurisation studies, and is therefore an important one to test here. The effect of variations in the initial pore pressure will be tested further below.

  In Equations \eqref{eq:dpdt} and \eqref{eq:dTdt}, all the fluid and rock properties are potentially varying (sometimes strongly) with pore pressure and temperature. Although well-chosen constant values can be used to determine typical estimates for the effect of thermal pressurisation \citep[e.g.,][]{rice06,rempel06,brantut16b}, here we aim to provide more accurate predictions and therefore we solve \eqref{eq:dpdt} and \eqref{eq:dTdt} including the full dependencies of parameters with pressure and temperature conditions.

  The coupled system (\ref{eq:dpdt}--\ref{eq:dTdt}) is solved using a semi-implicit finite difference method (similar to that used by \citet{brantut10}). The full list of parameter values is given in Table \ref{tab:parameters}, using the laboratory-derived permeability for each rock (intact/healed, and damaged tonalite).

  \begin{table}
    \caption{List of parameters used in the simulations.}
    \label{tab:parameters}
    \begin{center}
      \begin{tabular}{ll}
        \hline
        Parameter & Value\\
        \hline
        Normal stress, $\sigma_\mathrm{n}$ & $280$~MPa\\
        Initial pore pressure, $p_0$ & $150$~MPa\\
        Initial temperature, $T_0$ & $300^\circ$C\\
        Shear zone width, $h$ & $10$~$\mu$m\\
        Slip rate\tablenotemark{a}, $V$ &  --- \\
        Friction coefficient, $f$ & $0.15$\\
        Porosity, $n$ & $2$~\%\\
        Permeability\tablenotemark{b}, $k$ & $k_0e^{-\kappa P_\mathrm{eff}}$\\
        Specific heat, $\rho c$ & $2.7\times 10^{6}$~Pa~$^\circ$C$^{-1}$\\
        Thermal diffusivity, $\alpha_\mathrm{th}$ & $0.7\times10^{-6}$~m$^2$~s$^{-1}$\\
        Pore compressibility, $\beta_\mathrm{n}$ & $2\times 10^{-9}$~Pa$^{-1}$\\
        Pore thermal expansivity, $\lambda_\mathrm{n}$ & $0$~K$^{-1}$\\
        Fluid density\tablenotemark{c}, $\rho_\mathrm{f}$ & ---\\
        Fluid compressibility\tablenotemark{c}, $\beta_\mathrm{f}$ & ---\\
        Fluid thermal expansivity\tablenotemark{c}, $\lambda_\mathrm{f}$ & ---\\
        Fluid viscosity\tablenotemark{c}, $\eta$ & ---\\
        \hline
      \end{tabular}
      \small
      \tablenotetext{a}{Slip rate is given by a regularised Yoffe function with a maximum displacement of $0.2$~m, a total duration of $1.02$~s and a smoothing time of $0.01$~s.}
      \tablenotetext{b}{Permeability is given as a function of effective pressure $P_\mathrm{eff} = \sigma_\mathrm{n}-p$ using the nominal $k_0$ and pressure dependency $\kappa$ parameter determined in the laboratory.}
      \tablenotetext{c}{All water properties are given as a function of pore pressure and temperature using the International Association for the Properties of Water and Steam formulation \protect\citep{junglas09}.}
    \end{center}
  \end{table}

  \subsection{Results}

\begin{figure*}
    \centering
    \includegraphics{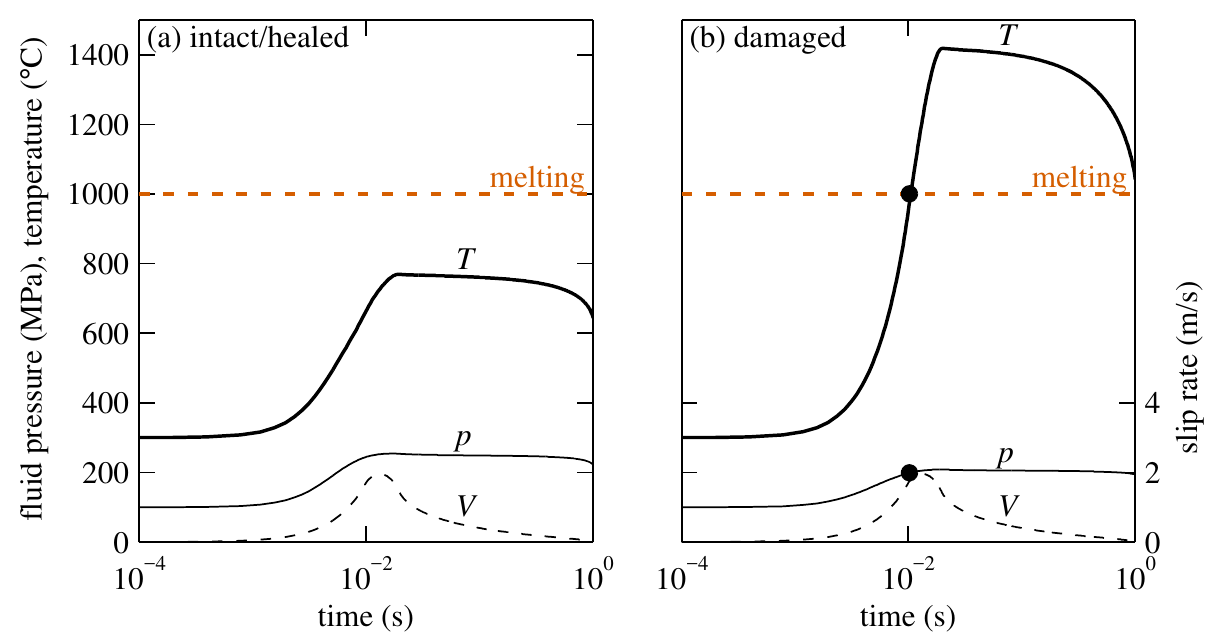}
    \caption{Reference computations of pore pressure ($p$) and temperature ($T$) with the slip rate history $V(t)$ using parameter values for (a) the intact (or damaged-healed) tonalite, and (b) the damaged tonalite.}
    \label{fig:endmembers}
  \end{figure*}

  We first performed reference computations using the parameter values as listed in Table \ref{tab:parameters}. The results are shown in Figure \ref{fig:endmembers}. In the intact rock, the peak temperature reached during slip remains relatively low, around $770^\circ$C, well below the bulk melting temperature of the tonalite. By contrast, in the damaged rock, the temperature rises beyond $1000^\circ$C after only $1$~cm slip, therefore predicting bulk melting early during the slip event. 
    
  Our simulations show that the presence of pseudotachylytes is not well explained if we use our laboratory-derived values of permeability for the rock directly hosting them in the field. According to our analysis, at initially hydrostatic fluid pressure conditions in the rupture zone, pseudotachylytes should only form in the damaged tonalite. In order to investigate what level of damage is required to start forming pseudotachylyte in an intact tonalite, we computed the peak temperature achieved during slip for a range of nominal permeabilities $k_0$ and pore space compressibilities $\beta_\mathrm{n}$. The compressibility of the pore space is poorly constrained, and its value is expected to be significantly affected by the occurrence of damage and microfracturing off the slip zone \citep[][Appendix A]{rice06}. 

\begin{figure*}
  \centering
  \includegraphics{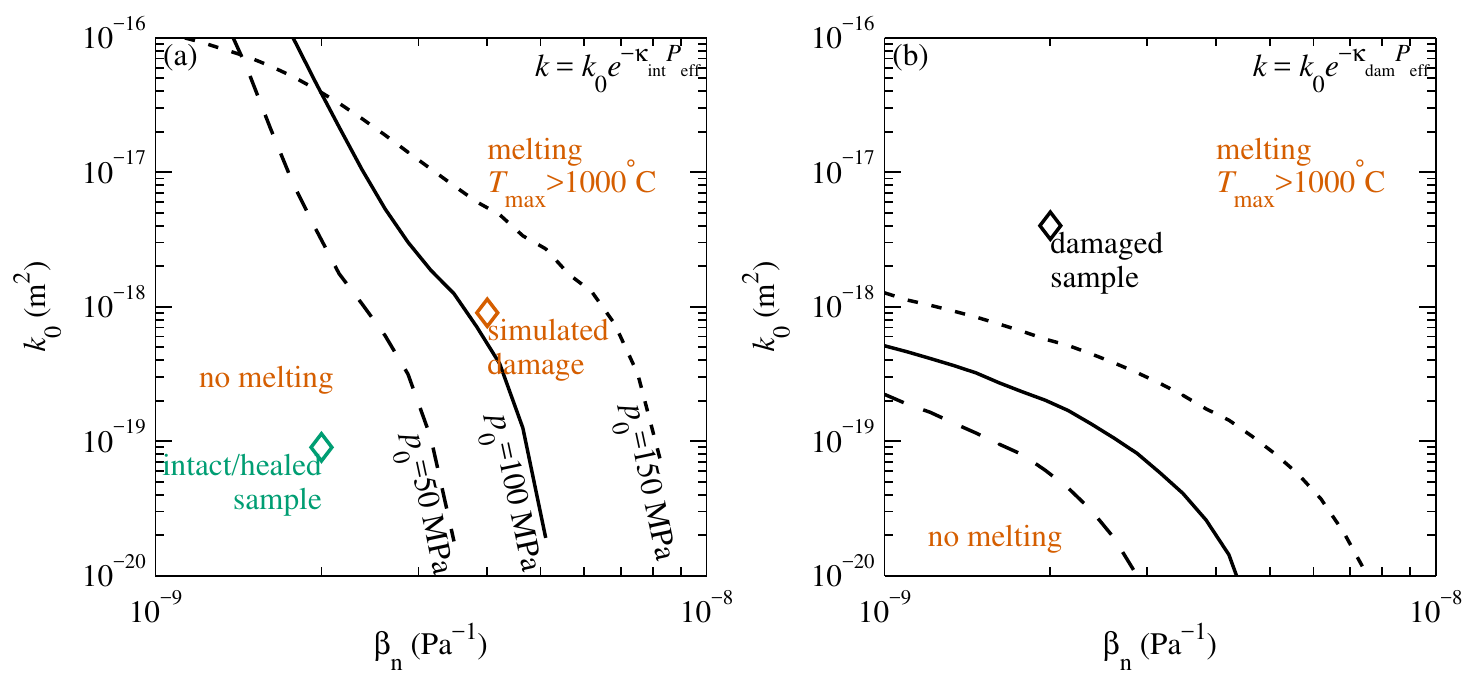}
  \caption{Contours showing the onset of melting (isotherm $1000^\circ$C) during seismic slip using the same slip rate history as in Figure \ref{fig:endmembers}, as a function of nominal permeability $k_0$ and pore space compressibility $\beta_\mathrm{n}$. The left and right panels are for simulations using the pressure sensitivity $\kappa$ of the intact (a) and damaged tonalite (b), respectively. Solid lines correspond to an initial hydrostatic pore pressure of $p_0=100$~MPa, dashed lines correspond to a lower initial pore pressure of $p_0=50$~MPa, simulating initial dilatancy, and dotted lines correspond to $p_0=150$~MPa, simulating supra-hydrostatic pore pressure conditions. In panel (a), the orange symbol labelled ``simulated damage'' corresponds to parameter values where $\beta_\mathrm{n}$ was doubled and $k_0$ was increased by a factor 10 compared to the intact values.}
  \label{fig:meltonset}
\end{figure*}

Figure \ref{fig:meltonset} reports the results of our computations, outlining the $1000^\circ$C isotherm as the onset of melting for each set of parameters. Starting from the intact or damaged-healed tonalite (Figure \ref{fig:meltonset}(a)), using an initial pore pressure $p_0=100$~MPa, we observe that a rather small increase in $\beta_\mathrm{n}$ up to around $5\times 10^{-9}$~Pa$^{-1}$ is sufficient to explain bulk melting of the rock.  By contrast, only very large increases in $k_0$ of around four orders of magnitude allow the system to reach a peak temperature above $1000^\circ$C. Overall, simply doubling $\beta_\mathrm{n}$ to around $4\times 10^{-9}$~Pa$^{-1}$ and increasing $k_0$ by one order of magnitude appears sufficient to explain bulk melting. A reduction in the initial pore pressure (dashed line), simulating instantaneous dilatancy at the onset of slip \citep{rice06}, tends to reduce the increase in $\beta_\mathrm{n}$ required to observe bulk melting to only a factor of 1.5. Conversely, an initially higher pore pressure level (dotted line) tends to limit the peak temperature achieved, and a larger increase in $\beta_\mathrm{n}$, by at least a factor of 4, would be required for melting to occur. 

Figure \ref{fig:meltonset}(b) shows the results obtained from varying $k_0$ and $\beta_\mathrm{n}$ but using the pressure sensitivity coefficient of permeability $\kappa_\mathrm{dam}$ of the damaged rock (which we recall is smaller than that of the intact rock). Compared to the case with $\kappa_\mathrm{int}$, changing $k_0$ has a stronger effect on the peak temperature; but around one order magnitude change is still required to significantly affect the onset of melting, while only a modest change in $\beta_\mathrm{n}$ produces dramatic variations in peak temperature, resulting in a strong control on the onset of melting.

\section{Discussion and Conclusions}
\label{sec:discussion}



%

  Our results constitute the first attempt to assess theoretical predictions of thermal pressurisation, using pseudotachylytes as a natural thermometer. Despite a rather conservative choice of parameters, the predictions of peak temperature based on laboratory-derived properties of the host rock (intact or damaged-healed) are inconsistent with the presence of pseudotachylytes. Instead, using the properties of damaged, unhealed rocks found further away from the fault leads to more consistent predictions of early bulk melting.

  In order to reconcile the field observations with the model prediction, one needs to increase the pore compressibility of the host rock (intact or damaged-healed) by a factor of 2 to 4. Equivalently, doubling the pore space compressibility and increasing the permeability by one order of magnitude will also allow reconciliation of the temperature predictions with the presence of frictional melt. These ad hoc modifications of rock properties to simulate damage were originally used by \citet{rice06}, and our results appear to validate this approach.

  Here we assumed that free fluids were present prior to seismic slip. This assumption may not be satisfied, in which case the presence of pseudotachylytes would be a natural consequence of frictional heating. However, what our results demonstrate is that the presence of pseudotachylytes is not incompatible with the occurrence of thermal pressurisation, once the effect of damage (either in terms of permeability, compressibility or dilatancy variations) have been accounted for. Dynamic fault weakening would then proceed by a sequence of mechanisms, from an initial stage where flash heating could be the main weakening process, followed by thermal pressurisation, and finally frictional melting.

  The common practice when applying the principle of thermal pressurisation to produce strength or temperature predictions is to use laboratory-derived rock properties on exhumed materials, neglecting dilatancy, and assuming an ambient hydrostatic pore pressure \citep[e.g.,][, among many others]{wibberley05,noda05,rice06}. What we show here is that all these assumptions need to be used with great caution, notably because \textit{in situ} rock properties and initial pore pressure conditions are likely strongly modified by the earthquake process itself \citep[see also][]{griffith12}, or by exhumation processes (e.g., cooling microfractures, although these can be easily closed by confining pressure). We also did not account for the additional sources of fluids associated with devolatilisation of hydrous phases due to frictional heating: Such reactions tend to buffer fault core temperature even more efficiently than thermal pressurisation \citep{sulem09,brantut10,brantut11}, so that our modelled temperatures should be viewed as upper bounds, reinforcing the conclusion that rock properties and initial conditions must have been significantly affected by fracturing processes.



The dominant parameter controlling the efficiency of thermal pressurisation is clearly the pore space compressibility, $\beta_\mathrm{n}$, and is also the most difficult to constrain experimentally. Another key factor is the permeability; however, while permeability only influences the hydraulic diffusivity of the rock, the pore space compressibility enters in both the hydraulic diffusivity and the thermal pressurisation coefficient. An increase in $\beta_\mathrm{n}$ has two competing effects: (1) It tends to reduce the hydraulic diffusivity, which would promote undrained pressurisation and reduce frictional heating, and (2) reduces the thermal pressurisation coefficient, therefore decreasing the efficiency of the thermal pressurisation mechanism. It appears that the latter effect largely dominates, and therefore even small changes in $\beta_\mathrm{n}$, for instance, due to microcrack damage, have a large influence on the effect of thermal pressurisation.

In addition, the overall low average hydraulic diffusivity, of the order of $10^{-6}$~m$^2$~s$^{-1}$, and the short duration of seismic events, of the order of a few seconds, implies that only the first few millimetres of rock around the slipping zone are controlling the overall behaviour due to thermal pressurisation. Hence, very local variations in damage have large scale consequences on the dynamic rheology of the fault.

A striking consequence is that relatively minor variations in microcrack damage can act as a toggle switch for the onset of bulk frictional melting. In the geological record, this effect is illustrated by the patchiness and sometimes erratic occurrence of pseudotachylytes along the length of a single fault strand \citep{smith13}. Indeed, along the GLFZ, \citet{griffith12} showed that the amount of microscale damage varies significantly along strike, together with the occurrence of pseudotachylytes. Such variations are expected to have first-order consequences on the rheology of the fault and therefore on the dynamics of ruptures: Indeed, the onset of melting is typically associated with a transient strengthening \citep[e.g.,][]{hirose05} and could promote pulse-like ruptures.

The fault rocks investigated here have been extracted from a relatively immature fault, in a crystalline basement, with overall small cumulative offsets \citep{ditoro04,smith13}. The nature and impact of microcrack damage in such rocks is likely to be quite different to that in mature faults containing finely comminuted, sometimes clay-rich fault cores. The absence of reported pseudotachylytes along these mature faults might be an indicator that either mature fault rocks are less prone to damage (being already formed by comminution processes), or that damage has a relatively minor impact on the temperature achieved during seismic slip.

One key aspect of damaged fault rocks is the degree of healing or sealing of the microcracks. The geological record along the Gole Larghe Fault shows that even healed or sealed fault rocks are hosts to pseudotachylytes, and our simulation results confirm that at least some damage (in the form of an increased compressibility and permeability) must have occurred transiently to allow bulk frictional melting. Therefore, it appears that long-term damage recovery (due to healing or sealing) may be rapidly overprinted by newly generated damage during an earthquake. This implies that recovery processes may only impact the early stages of seismic slip, the later stages being largely dominated by the transient effects of newly generated damage.

\begin{acknowledgments}
  This work was partially supported by the UK Natural Environment Research Council through grants NE/K009656/1 to NB and NE/M004716/1 to TMM and NB. We thank Giulio Di Toro for his support and for providing us with the opportunity to perform field work in the Adamello massif. Steve Cox and Marie Violay provided critical and helpful review of this manuscript. The data used in this paper are available as supporting information or upon request to the corresponding author.
\end{acknowledgments}


\end{article}

\end{document}